\documentclass[aps,prl,twocolumn,amsmath,amssymb,superscriptaddress,citeautoscript]{revtex4-2}

\usepackage{graphicx}
\usepackage{natbib}
\usepackage{dcolumn}
\usepackage{bm}
\usepackage{color}
\usepackage{hyperref}

\newcommand{\MN}[2]{{\color{black}#2}}                

\begin{document}


\title{Electron doping of the iron-arsenide superconductor CeFeAsO\\ controlled by hydrostatic pressure }

\author{K.\ Mydeen}
\affiliation{Max-Planck Institute for Chemical Physics of Solids, N\"{o}thnitzer Str.\ 40, 01187 Dresden}

\author{A.\ Jesche}
\affiliation{Max-Planck Institute for Chemical Physics of Solids, N\"{o}thnitzer Str.\ 40, 01187 Dresden}
\affiliation{Experimental Physics VI, Center for Electronic Correlations and Magnetism, Institute of Physics, University of Augsburg, 86135 Augsburg, Germany}

\author{K.\ Meier-Kirchner}
\affiliation{Max-Planck Institute for Chemical Physics of Solids, N\"{o}thnitzer Str.\ 40, 01187 Dresden}

\author{U.\ Schwarz}
\affiliation{Max-Planck Institute for Chemical Physics of Solids, N\"{o}thnitzer Str.\ 40, 01187 Dresden}

\author{C.\ Geibel}
\affiliation{Max-Planck Institute for Chemical Physics of Solids, N\"{o}thnitzer Str.\ 40, 01187 Dresden}

\author{H.\ Rosner}
\affiliation{Max-Planck Institute for Chemical Physics of Solids, N\"{o}thnitzer Str.\ 40, 01187 Dresden}

\author{M.~Nicklas}
\email{Michael.Nicklas@cpfs.mpg.de}
\affiliation{Max-Planck Institute for Chemical Physics of Solids, N\"{o}thnitzer Str.\ 40, 01187 Dresden}

\begin{abstract}
In the iron-pnictide material CeFeAsO not only the Fe moments, but also the local 4$f$ moments of the Ce order antiferromagnetically at low temperatures. We elucidate on the peculiar role of the Ce on the emergence of superconductivity.  While application of pressure suppresses the iron SDW ordering temperature monotonously up to 4~GPa, the Ce-4$f$ magnetism is stabilized, until both types of magnetic orders disappear abruptly and a narrow SC dome develops. With further increasing pressure characteristics of a Kondo-lattice system become more and more apparent in the electrical resistivity.  This suggests a connection of the emergence of superconductivity with the extinction of the magnetic order and the onset of Kondo-screening of the Ce-4$f$ moments.

\end{abstract}

\date{\today}

\maketitle


Unconventional superconductors, such as the heavy-fermion \cite{Steglich1979}, organic \cite{Jerome1980}, high-$T_c$ cuprate \cite{Muller1986}, and iron-based superconductors \cite{Hosono2008}, belong to diverse material classes. On a very first glance these classes do not have a lot in common except that the superconductivity develops in the proximity to a magnetically ordered phase. This provides a phenomenological recipe to drive a material belonging to one of the families to superconductivity by suppressing the magnetic order as function of some external control parameter, such as charge doping, chemical substitution, or external pressure. The physical mechanism behind appears to be completely different and suggests the presence of a more general physical principle behind. CeFeAsO connects both, the physics of iron-based superconductors and of heavy-fermion metals. It is a parent compound to iron-pnictide superconductors and the corresponding phosphorous compound CeFePO is a heavy-fermion metal \cite{Jesche2009,Bruening2008}.

At ambient pressure, CeFeAsO exhibits a structural phase transition from tetragonal (T) to orthorhombic (O) around $T_0=150$~K. At a slightly lower temperature $T_N^{\rm Fe}$ the itinerant iron moments form a commensurate spin-density wave (SDW), and below $T_N^{\rm Ce} =3.7$~K the localized cerium $4f$ moments order antiferromagnetically \cite{Jesche2009,Jesche2010}. Application of external pressure suppresses the SDW ordering, but no superconductivity has been reported \cite{Zocco2011,Materne18}.
In contrast, superconductivity was found either by charge doping \cite{Chen2008,Zhao2010}, by inducing oxygen vacancies \cite{Ren2008}, by hydrogen substitution \cite{Iimura12,Matsuishi12}, or by isoelectronic substitution of As by P in CeFeAs$_{1-x}$P$_x$O \cite{Luo2010,Jesche2012,Mydeen2012}. The latter can be also considered as an application of chemical pressure. Appropriate P substitution in CeFeAsO not only suppresses the Fe-SDW ordering and leads to a narrow superconducting (SC) phase, but also changes the way the Ce moments order from antiferromagnetic (AFM) to ferromagnetic (FM). This change in the type of ordering of the rare-earth moments, which is unique among the \textit{R}FeAsO (\textit{R} = rare-earth element), takes place nearby the P-concentration, where superconductivity is observed.
On the P-rich side of the phase diagram, CeFePO is a heavy-fermion metal in close proximity to a FM instability \cite{Bruening2008,Holder2010,Lausberg12,Jesche2017}. This suggests a new and proper route of electron doping a stoichiometric iron-pnictide material by application of hydrostatic pressure and taking advantage of the Kondo effect. Due to the Kondo effect the Ce-$4f$ electrons become part of the Fermi-surface corresponding to an electron doping of CeFeAsO.


The details of the preparation and characterization of the CeFeAsO single crystals have been reported in Ref.\ \onlinecite{Jesche2009}.
The temperature dependence of the electrical resistivity under pressure was measured in a physical property measurement system (Quantum Design), utilizing a resistance bridge (LR700, Linear Research), in magnetic fields up to 9~T using a diamond-anvil cell (DAC). Magnetization data was recorded using a miniature DAC in a magnetic property measurement system (Quantum Design). The magnetization of the sample was obtained by measuring the DAC with sample and subsequently subtracting the cell background recorded in a separate run. In both pressure cells glycerin served as pressure transmitting medium. The pressure inside the DAC was determined by a standard ruby fluorescence method at room temperature  before and after each cooling down (for details on the pressure homogeneity see the Supplemental Material \cite{suppl}).
X-ray diffraction under pressure was conducted on ground crystalline powder from the same batch of CeFeAsO single crystals. The samples were loaded into a membrane-driven DAC. Helium served as pressure-transmitting medium. A helium gas-flow cryostat enabled controlled low-temperature measurements. Sm-doped SrB$_4$O$_7$ was used as pressure calibrant. No pressure gradient could be detected inside the pressure chamber. The data were collected on ID9A at the ESRF, Grenoble, using a wavelength of 41.44~pm. The recorded two-dimensional diffraction patterns were integrated by means of the computer program FIT2D.

\begin{figure}[t!]
\begin{center}
\includegraphics[clip,width=0.97\columnwidth]{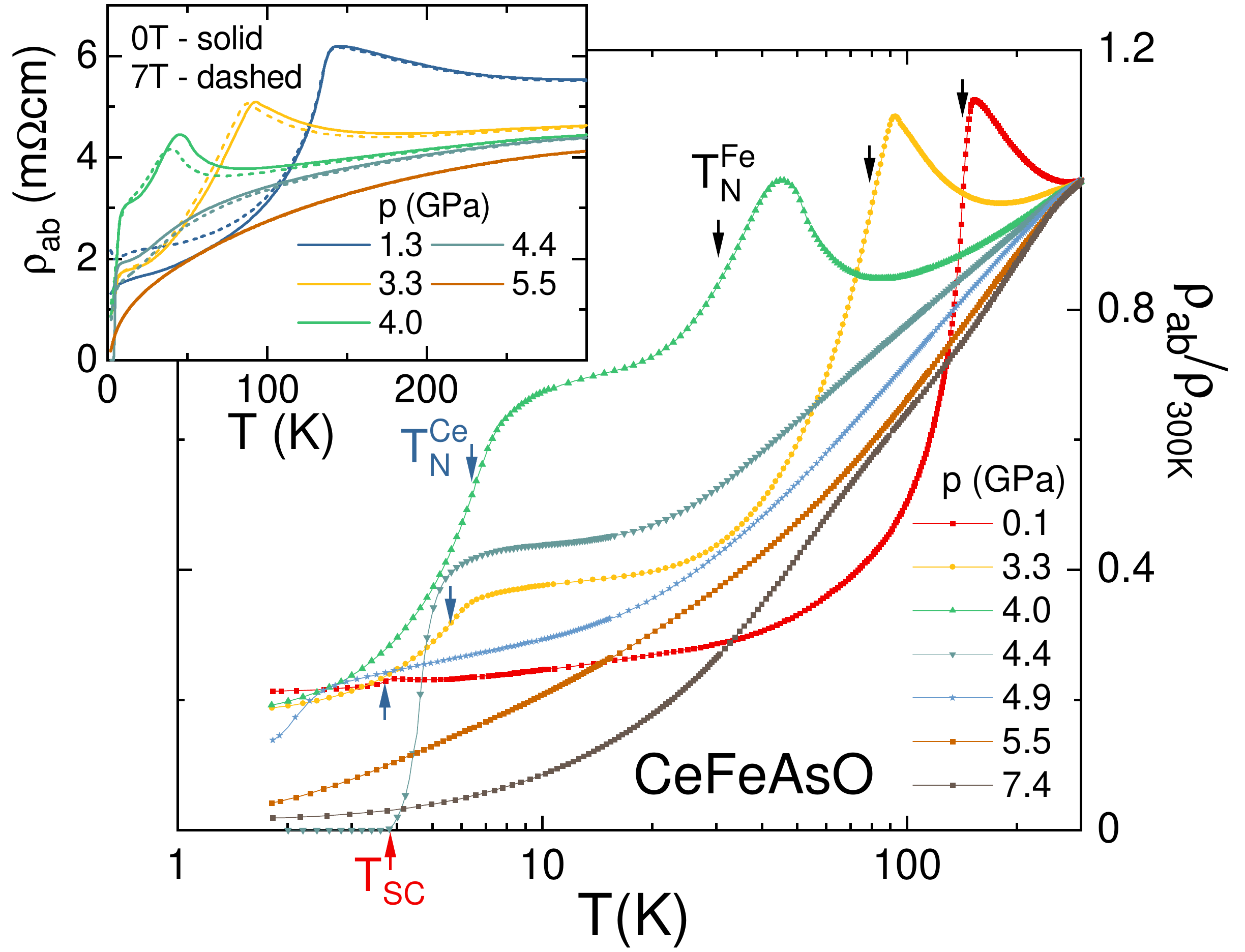}
  \caption{Temperature dependence of the in-plane resistivity $\rho_{ab}(T)$ normalized by its value at 300~K of a CeFeAsO single crystal for selected hydrostatic pressures up to 7.4~GPa on a logarithmic temperature scale. Black, blue, and red arrows mark the SDW ordering of the iron moments, the AFM ordering of the localized Ce moments, and the superconducting transition temperature, respectively. Inset: $\rho_{ab}(T)$ in 0 and 7~T magnetic field, $H\parallel c$, for selected pressure.
  \label{fig1}}
\end{center}
\end{figure}


At ambient pressure, CeFeAsO displays distinct anomalies in the in-plane electrical resistivity $\rho_{ab}(T)$ indicating the SDW ordering of the iron moments and the AFM ordering of the localized Ce moments (see Fig.\ \ref{fig1}). The SDW ordering is closely connected to a structural transition from a tetragonal to an orthorhombic phase. The magnetic transition temperatures $T_N^{\rm Fe}$ and $T_N^{\rm Ce}$ correspond to  maxima and the structural transition at $T_0$ to a shoulder in ${\rm d}\rho_{ab}(T)/{\rm d}T$. We follow the same analysis as in Ref.\ \onlinecite{Jesche2010}. The derivatives for selected pressures are shown in Fig.\ \ref{fig2}a (further details can be found in the Supplemental Material \cite{suppl}). We obtain $T_0=149$~K and $T_N^{\rm Fe} = 142$~K, and $T_N^{\rm Ce} = 3.7$~K at ambient pressure. $T_0$ from resistivity agrees well with our structural data. For details on the analysis of the structural data see the Supplemental Material \cite{suppl}. We note that all values are in good agreement with reports in literature \cite{Jesche2009,Jesche2010}.

The normalized in-plane electrical resistivity $\rho_{ab}(T)/\rho_{ab}({\rm 300K})$ of single crystalline CeFeAsO for selected hydrostatic pressures is depicted in Fig.\ \ref{fig1}.  Application of hydrostatic pressure has a strong effect on the shape of the resistivity curves in CeFeAsO. The pronounced maximum shifts to lower temperature and then disappears completely upon increasing pressure. At first, up to about 2~GPa $T_N^{\rm Fe}$ and $T_0$ exhibit only a weak pressure dependence. With further increasing pressure both transition temperatures start to decrease with a much larger rate but, simultaneously, also the splitting between them increases significantly. At $p = 4.0$~GPa, $\rho_{ab}(T)$ still exhibits a pronounced narrow maximum related to the structural transition and the SDW ordering, at $T_0\approx 49$~K and  $T_N^{\rm Fe} \approx  30$~K, respectively, before the signatures of both transitions are abruptly lost above 4.0~GPa. The sharp suppression of the Fe-SDW ordering has been seen in M\"{o}ssbauer data too \cite{Materne18}.

In contrast to the Fe-SDW ordering at high temperatures the AFM ordering temperature of the localized Ce-$4f$ moments increases with increasing pressure with a slope of roughly 1~{\rm K/GPa} in good agreement with Ref.\ \onlinecite{Zocco2011}. Up to 4~GPa we do not find any indication of a change in the type of magnetic ordering in contrast to the findings under the application of chemical pressure \cite{Luo2010,Jesche2012,Mydeen2012}. The signatures of $T_N^{\rm Fe}$ and $T_N^{\rm Ce}$ in the resistivity data disappear simultaneously in a step-like fashion manifesting the close connection between the $3d$-iron and $4f$-cerium magnetism in CeFeAsO \cite{Chi2008,Maeter2009}. At $p = 4.4$~GPa, where no long range magnetic order, neither of the Fe nor the Ce moments, is present anymore, superconductivity starts to develop at low temperatures. In contrast to the results of isoelectronic substitution of As by P in CeFeAs$_{1-x}$P$_x$O, superconductivity does not coexist with ferromagnetically ordered Ce moments \cite{Luo2010,Jesche2012}.

At ambient pressure the iron SDW order in CeFeAsO is extremely robust against application of magnetic field, but upon increasing pressure that changes and the Fe-SDW ordering becomes extremely sensitive to magnetic field. At zero pressure the transition temperature is not affected by a field of 40~T \cite{Yuan2011}. In contrast to that, at 4.0~GPa a field of only 7~T drives the transition anomaly in $\rho_{ab}(T)$ by about 6~K toward lower temperatures, while its shape remains nearly unchanged (see inset of Fig.\ \ref{fig1}). This behavior points at a weakening of the SDW ordering upon increasing pressure. It is remarkable that for all pressures the application of magnetic field leads to an almost rigid shift of the resistivity curves in the region of the SDW transition. Furthermore, upon cooling the magnetoresistance ${\rm MR_{\rm7T}}=[\rho_{\rm 7T}(T)-\rho_{\rm 0T}(T)]/\rho_{\rm 0T}(T)$ starts to increase already below $T_0$, above the Fe-SDW ordering at $T_N^{\rm Fe}$ (see Fig.\ \ref{fig2}b). The large negative magnetoresistance above $T_N^{\rm Fe}$ suggest the presence of strong magnetic fluctuations, which are suppressed by magnetic field.

\begin{figure}[tb!]
\begin{center}
  \includegraphics[clip,width=0.97\columnwidth]{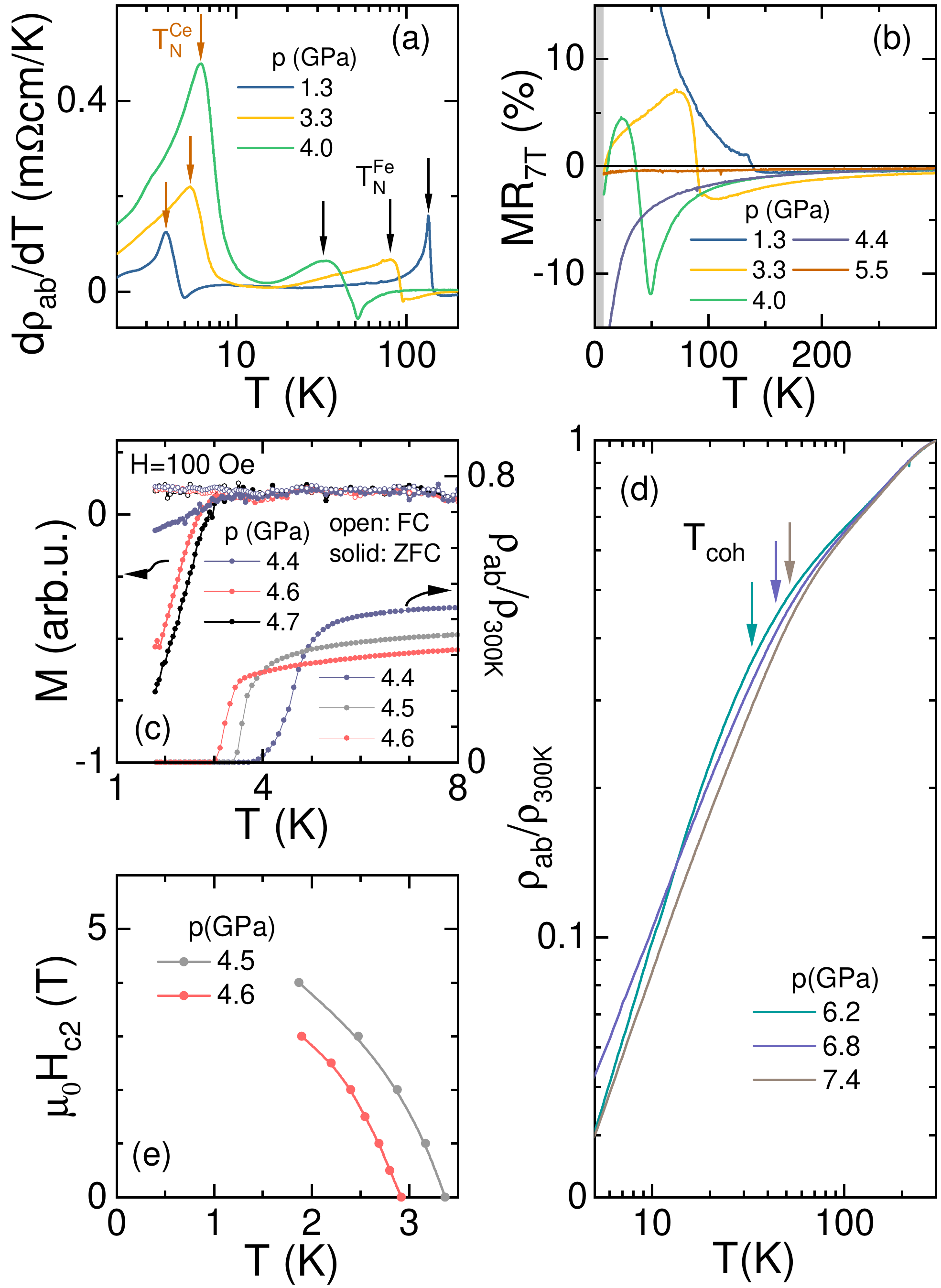}
  \caption{
  (a) Temperature derivative of the resistivity ${\rm d}\rho_{ab}(T)/{\rm d}T$ plotted on a logarithmic temperature scale. Arrows mark the position of  $T_N^{\rm Fe}$ and $T_N^{\rm Ce}$.
  (b) Magnetoresistance ${\rm MR_{\rm7T}}=[\rho_{\rm 7T}(T)-\rho_{\rm 0T}(T)]/\rho_{\rm 0T}(T)$ for different pressures.
  (c) Left axis: Temperature dependence of the field cooled (FC) and zero-field cooled (ZFC) magnetization data for selected pressures. Right axis: Selected $\rho_{ab}(T)/\rho_{\rm 300K}$ curves showing the SC transition.
  (d) $\rho_{ab}(T)/\rho_{\rm 300K}$ on a double-logarithmic representation. The Kondo-coherence temperature $T_{\rm coh}$ is indicated by arrows. See text for details.
  (e) Superconducting $H-T$ phase diagram of CeFeAsO at $4.5$ and 4.6 GPa.
  \label{fig2}}
\end{center}
\end{figure}

Once the magnetic order is removed from CeFeAsO, superconductivity appears suddenly in a narrow pressure range. The low temperature resistance in the pressure region of the SC phase is displayed in Fig.\ \ref{fig2}c. The SC transition at 4.4~GPa appears to be rather broad in $\rho_{ab}(T)$.  The onset is at $T_{{sc}{,\rm onset}}^{\rho} = 5.42$~K, while zero resistance is observed only below $T^{\rho}_{sc} = 3.61$~K. This corresponds to a width of $\Delta T_{sc}^{\rho} = 1.81$~K. With further increasing pressure the transition sharpens, $\Delta T_{sc}^{\rho} = 1.4$~K at 4.6~GPa, but $T_{sc}^{\rho}$ shifts toward lower temperatures. Temperature dependent magnetization $M(T)$ data, shown in  Fig.\ \ref{fig2}c, indicate the bulk nature of the superconductivity. At 4.4~GPa, a weak diamagnetic signal starts to develop below $T_{sc}^{\chi}$ in the zero-field cooled (ZFC) data. Upon further increasing pressure the diamagnetic signal becomes larger and $T_{sc}^{\chi}$ shifts toward lower temperatures. At 1.8~K, the lowest accessible temperature in our setup, we find a significant diamagnetic response at 4.6 and 4.7~GPa.
The absence of a signal in the field-cooled (FC) data can be explained by the presence of strong flux pinning which is commonly observed in iron-based superconductors. We note the good agreement between the onset of the diamagnetic response and the temperature below which a zero resistance is detected. Above 4.7~GPa no signature of superconductivity is visible in the $M(T)$ data anymore ($T > 1.8$~K).

At 4.5 and 4.6~GPa we have recorded $\rho_{ab}(T)$ in different magnetic fields to establish the SC $H-T$ phase diagram (see Fig.\ \ref{fig2}e and the Supplemental Material \cite{suppl}). Above 5~T we do not find any indication of superconductivity down to 1.8~K. The $T=0$ limit of the upper-critical field $\mu_0H_{c2}(0)|_{H\parallel c}$, for 4.6~GPa is smaller than that for 4.5~GPa. For both pressures $\mu_0H_{c2}(0)$ can be estimated to be between 4 and 5.5~T. This is significantly smaller than the orbital limiting field $H_{c2}^{\rm orb}(0)$. For the dirty limit it can be estimated by $H_{c2}^{\rm orb}(0)=0.69|{\rm d}H_{c2}/{\rm d}T|\times T_{sc}$ to be 11.6 and 8.8~T for 4.5 and 4.6~GPa, respectively \cite{Helfand1966}, indicating that Pauli spin-paramagnetic effects play an important role as pair-breaking mechanism.

The shape of the resistivity curves changes qualitatively between 4.0 and 4.4~GPa, as can be seen in Fig.\ \ref{fig1}, indicating drastic differences in the scattering processes at both pressures. The shape of $\rho_{ab}(T)$ at 4.0~GPa is marked by magnetic scattering processes and a clear feature at the Fe-SDW/structural transition. At 4.4~GPa \MN{}{and above} no indication for any magnetic ordering remains in $\rho_{ab}(T)$. Only the negative ${\rm MR_{\rm7T}}$ (Fig.\ \ref{fig2}b), below 30~K still reveals the presence of magnetic fluctuations at this pressure: the applied field quenches the magnetic fluctuations and removes in that way a scattering channel for the charge carriers leading to a reduced resistivity in magnetic field. The negative ${\rm MR_{\rm7T}}$ remains present above the SC transition temperature upon increasing pressure. Once the SC phase has disappeared at higher pressures, magnetic fields up to 7~T do not show any visible effect on $\rho_{ab}(T)$ anymore (see Fig.\ \ref{fig2}b). Therefore, we can conclude the absence of \MN{}{ any magnetic ordering and} magnetic fluctuations in the high pressure regime ($p \gtrsim 5$~GPa).

\begin{figure}[t!]
\begin{center}
  \includegraphics[clip,width=0.97\columnwidth]{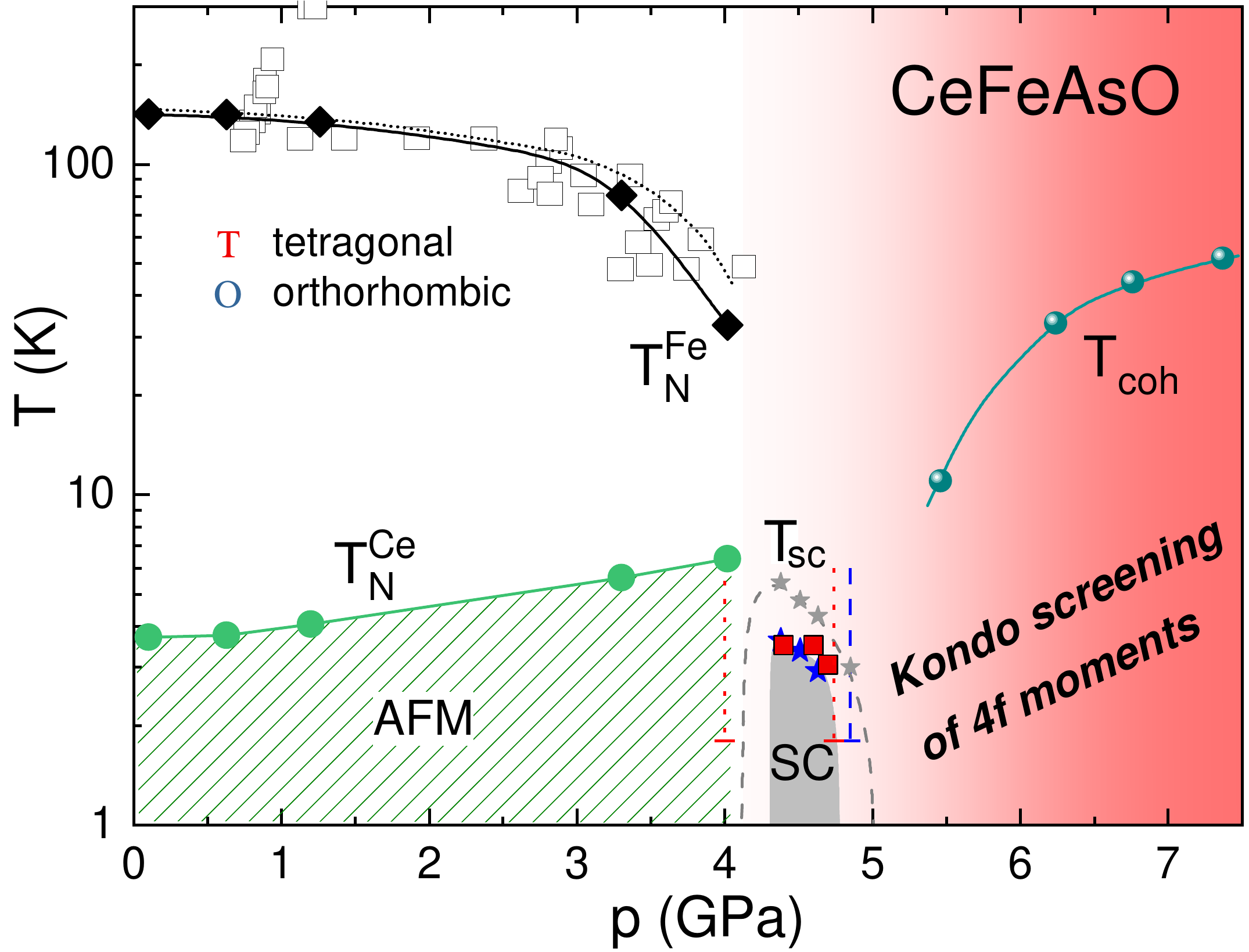}
  \caption{Pressure dependence of the Fe-SDW ordering temperature, $T_N^{\rm Fe}$ (filled black diamonds), structural phase transition $T_0$ from tetragonal (T) to orthorhombic (O) from high pressure x-ray diffraction (dotted line), AFM ordering temperature of Ce moments $T_N^{\rm Ce}$ (filled green circles), the SC transition temperature $T_{sc}$ observed from the onset of the diamagnetic signal (filled red squares) and from the onset (filled gray stars) and final (filled blue stars) resistive transition. The lines are guides to the eye. The vertical dotted red and dashed blue lines indicate pressures where magnetization and resistivity experiments have been carried out but do not detect signatures of superconductivity. $T_{\rm coh}$ is the Kondo-coherence temperature.
  \label{fig3}}
\end{center}
\end{figure}


The $T-p$ phase diagram of CeFeAsO, shown in Fig.\ \ref{fig3}, is exceptional among the iron-pnictide superconductors: upon increasing pressure the iron SDW magnetism becomes highly susceptible to the application of a magnetic field, $T_N^{\rm Fe}$ and $T_N^{\rm Ce}$ disappear abruptly, and superconductivity develops only in a narrow SC dome. The observation of a narrow SC dome is in strong contrast to observations on charge doping by elemental substitution where a broad SC region is reported, \textit{i.e.}\ hydrogen anion substitution studies evidence the existence of a broad SC dome \cite{Iimura12,Matsuishi12}. In particular, there superconductivity extends to regions far away from the critical point for the suppression of the iron SDW phase. Our phase diagram is much closer to the theoretical predictions for spin-fluctuation mediated superconductivity (\textit{e.g.}\ Ref.\ \onlinecite{Shiliang2011}). Here, a narrow superconducting dome is expected around the critical pressure. We indeed observe a narrow, but not symmetric SC dome.

Superconductivity is only established once the local moment Ce-$4f$ order is suppressed. The reason for this behavior may be originating in the peculiar role of the Ce-$4f$ moments.
We note that our data \MN{}{$p\geq 6.2$}~GPa follow a $\rho=\rho_0+AT^2$ dependence at low temperatures indicative of a Fermi-liquid groundstate \cite{Landau1957}. The $A$-coefficient increases upon increasing pressure, \textit{i.e.}\ enlarging the distance to the SC phase (see Supplemental Material \cite{suppl}). These results suggest that quantum criticality of Ce moments is not the source of the
appearance of pressure-induced superconductivity in CeFeAsO. Furthermore, the abrupt disappearance
of the local moment cerium ordering at the critical pressure is in contrast to an AFM quantum critical point scenario, in which the SWD order of Ce-$4f$ moments is suppressed continuously to zero temperature (see \textit{e.g.}\ Ref.\ \onlinecite{Stewart2001}).

It has been shown that the Ce-$4f$ and the Fe-$3d$ magnetism are closely related in CeFeAsO \cite{Chi2008,Maeter2009,Zhang13}. Therefore, a change in the character of the Ce-$4f$ magnetism is expected to have a strong influence on the Fe-$3d$ magnetism and the SC properties. Indeed the temperature dependence of the resistivity evidences such a change upon increasing pressure.
At low pressures the well-localized moments of the Ce-$4f$ electrons order antiferromagnetically, marked by a distinct drop in the resistivity, while at high pressures, above the SC dome, the shape of $\rho_{ab}(T)$  is reminiscent of a Kondo-lattice system  (see \MN{}{Figs.\ \ref{fig1} and \ref{fig2}d}) \cite{Doniach1977,Grewe1991,Wirth2016,Thompson94}. In a Kondo-lattice system a maximum or shoulder in the measured resistivity corresponds to a maximum in the $4f$ contribution $\rho_{4f}(T)$ to the resistivity. This maximum indicates the onset of Kondo coherence. In a dense array of localized moments, \textit{i.e.}\ a Kondo-lattice, first a logarithmic increase in $\rho_{4f}(T)$ on lowering temperatures becomes apparent before coherence effects lead to a strong reduction in the $4f$ contribution to the resistivity upon further decreasing the temperature \cite{Grewe1991,Thompson94}. This behavior we observe in our data, $\rho_{ab}(T)$ decreases weakly at high temperature, exhibits a shoulder and decreases more strongly toward low temperatures as seen in Fig.\ \ref{fig2}d. We define the coherence temperature $T_{\rm coh}$ as the temperature where the resistivity starts to drop more rapidly (see the Supplemental Material \cite{suppl} for details on the determination of $T_{\rm coh}$).

In Ce-based Kondo-lattice systems $T_{\rm coh}(p)$ exhibits generally a positive pressure dependence \cite{Schilling79,Thompson1987,Thompson94} as also observed here (see Figs.\ \ref{fig2}d and \ref{fig3}). In CeFeAsO $T_{\rm coh}(p)$ can be extrapolated to zero temperature in the pressure region where the magnetic ordering disappears and superconductivity develops. This suggests an essential role of Kondo physics and, in particular, the screening of the Ce-$4f$ moments, for the emergence of superconductivity in CeFeAsO under hydrostatic pressure. This is in contrast to the observation of the effect of chemical pressure by phosphorous substitution in CeFeAs$_{1-x}$P$_x$O, where the FM ordering of local Ce-$4f$ moments prevents the formation of a robust superconducting phase \cite{Luo2010,Jesche2012}. \MN{}{We note that the Kondo-lattice scenario explains in a natural way the absence of local moment Ce-$4f$ magnetism in CeFeAsO at high pressures.}
Due to the increasing hybridization of conduction and Ce-$4f$ electrons upon increasing pressure \cite{Thompson94}, the Ce-$4f$ electrons become part of the Fermi surface \cite{Yamanaka97,Oshikawa2000}. In that way the Kondo effect, leading to a screening of the localized $4f$ moments by the conduction electrons, provides an effective mechanism of electron doping in CeFeAsO controlled by hydrostatic pressure. \MN{}{Even though our data provide strong evidence for a Kondo-lattice scenario at high pressures in CeFeAsO, further studies would be desirable, but are highly challenging in the high pressure regime.}

The maximum $T_{sc}\approx 5.4$~K found in the current pressure study is significantly lower than the ones found in F-doped CeFeAsO \cite{Chen2008}; the increase of the SC critical temperature from $26$~K for 8\%-F to $41$~K for 16\%-F indicates a relation between $T_{sc}$ and the charge carrier concentration. Accordingly, the comparatively low $T_{sc}$ under pressure in CeFeAsO could be interpreted as a weak effective charge doping caused by slight rearrangements of the electronic band structure. However, groundstate and maximum $T_{sc}$ in CeFeAsO are determined by a complex interplay of iron SDW, cerium AFM/FM, and Kondo interactions and their competition with superconductivity. In Co-doped CeFeAsO, for example, maximum $T_{sc}$'s of somewhat above 10~K were observed \cite{Prakash2009,Shang2013,Prando2013}. Isoelectronic P-doping revealed a maximum $T_{sc}$ of 4~K \cite{Jesche2012}, which is very similar to that found in the present work, despite the nonequivalence of hydrostatic and chemical pressure \cite{Materne18}.


In summary, CeFeAsO is an exceptional material among the iron-pnictide superconductors and provides a bridge to the heavy-fermion metals. On one hand it displays the generic phase diagram of the iron-pnictide superconductors, external pressure suppresses an SDW order and a superconducting dome develops in the vicinity of the critical pressure where the magnetic order disappears. On the other hand, the low temperature physics is closely connected to the Ce-$4f$ electrons. The emergence of the Kondo-effect and the screening of Ce-$4f$ moments by the conduction electrons lead to an effective electron-doping of CeFeAsO and seem to be essential for the development of superconductivity under external pressure.

We acknowledge Michael Hanfland for discussions and technical support during the experiment HC~1471 at the ESRF Grenoble. J.E.\ acknowledges support by the Deutsche Forschungsgemeinschaft (DFG, German Research Foundation) Grant No.\ JE748/1.

\bibliography{CeFeAsO}

\clearpage

\begin{center}
\textbf{\large Supplemental Material: Electron doping of the iron-arsenide superconductor CeFeAsO controlled by hydrostatic pressure}
\end{center}
\setcounter{equation}{0}
\setcounter{figure}{0}
\setcounter{table}{0}
\setcounter{page}{1}
\makeatletter

\renewcommand{\figurename}{Figure S\!\!}

\section*{X-ray diffraction experiments}

\begin{figure}[b!]
\includegraphics[width=0.85\linewidth]{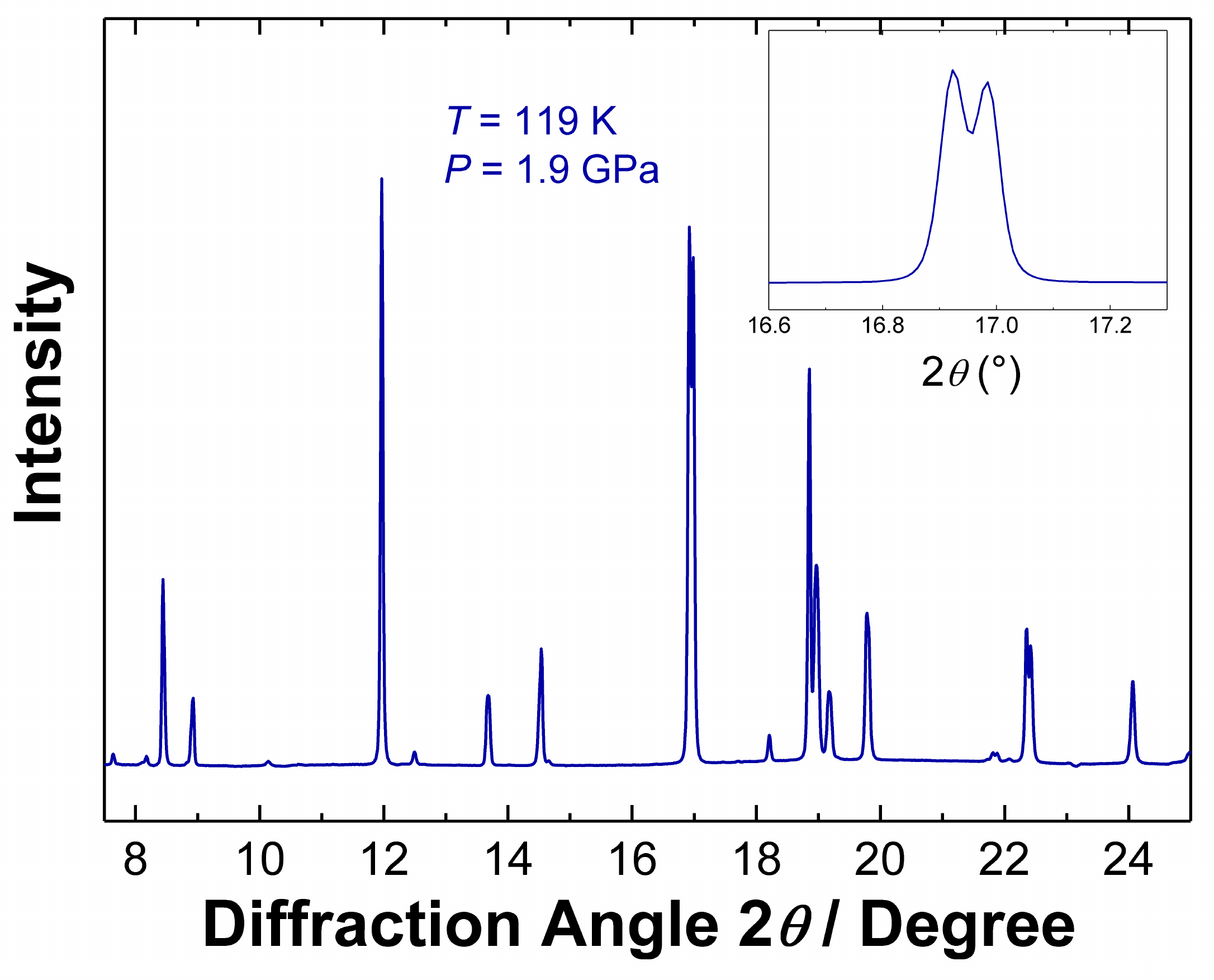}
\includegraphics[width=0.85\linewidth]{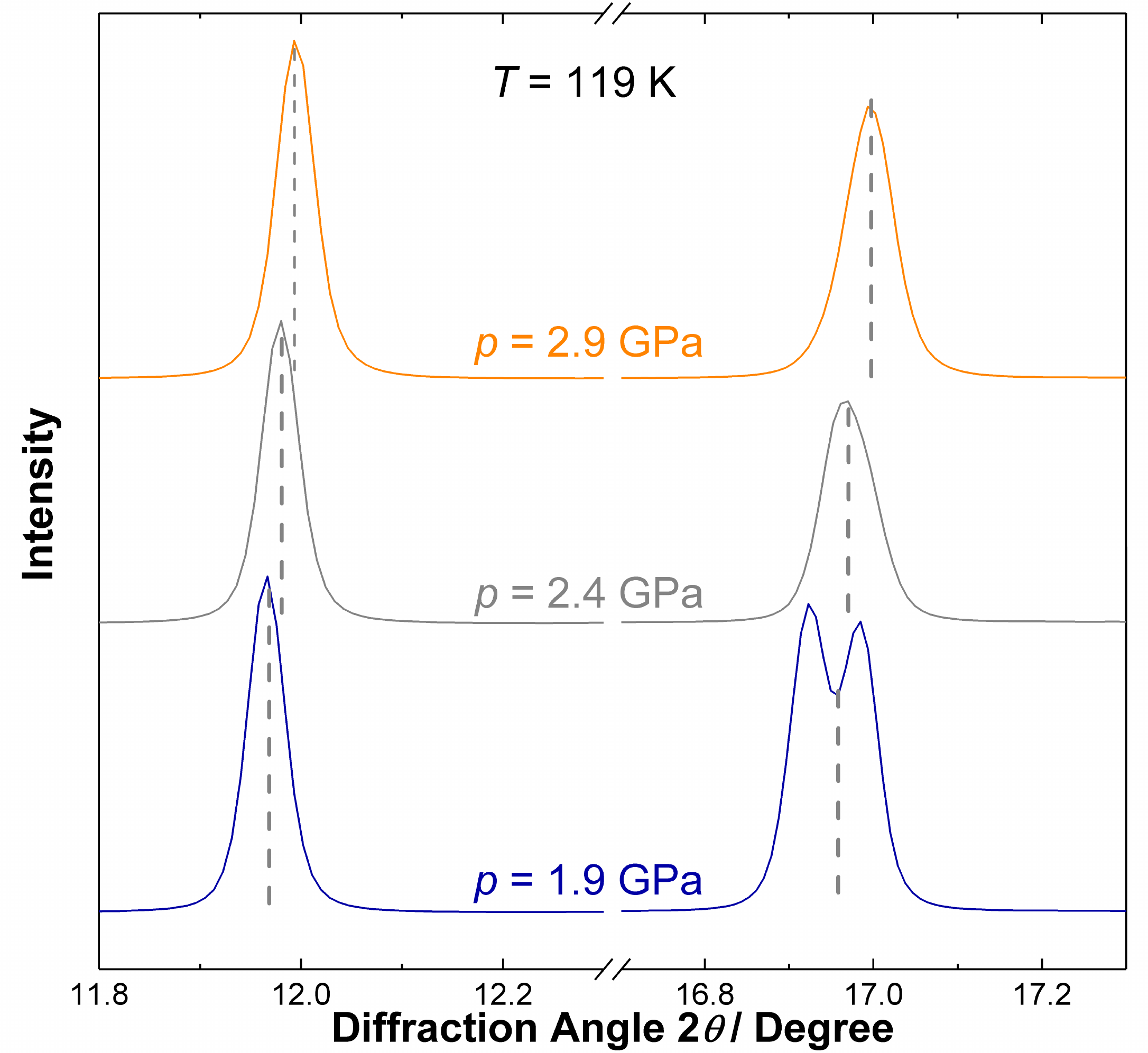}
\centering
\caption{Upper panel: Typical x-ray diffraction pattern for CeFeAsO ($T=119$~K, $p=1.9$~GPa) using synchrotron radiation and a diamond anvil cell as pressure apparatus. The inset shows a zoom for a peak (at about 17$^\circ$) that is split in the orthorhombic phase.
Lower panel: X-ray diffraction pattern for CeFeAsO  for an isothermal ($T=119$~K) pressure run ($p=1.9$~GPa, 2.4~GPa, and 2.9~GPa, from bottom to top). The peak at about 17$^\circ$ splits in the orthorhombic phase, whereas the other at about 12$^\circ$ does not split and serves as a reference peak.
\label{diff} }
\end{figure}
\begin{figure}[htb!]
\includegraphics[width=1.0\linewidth]{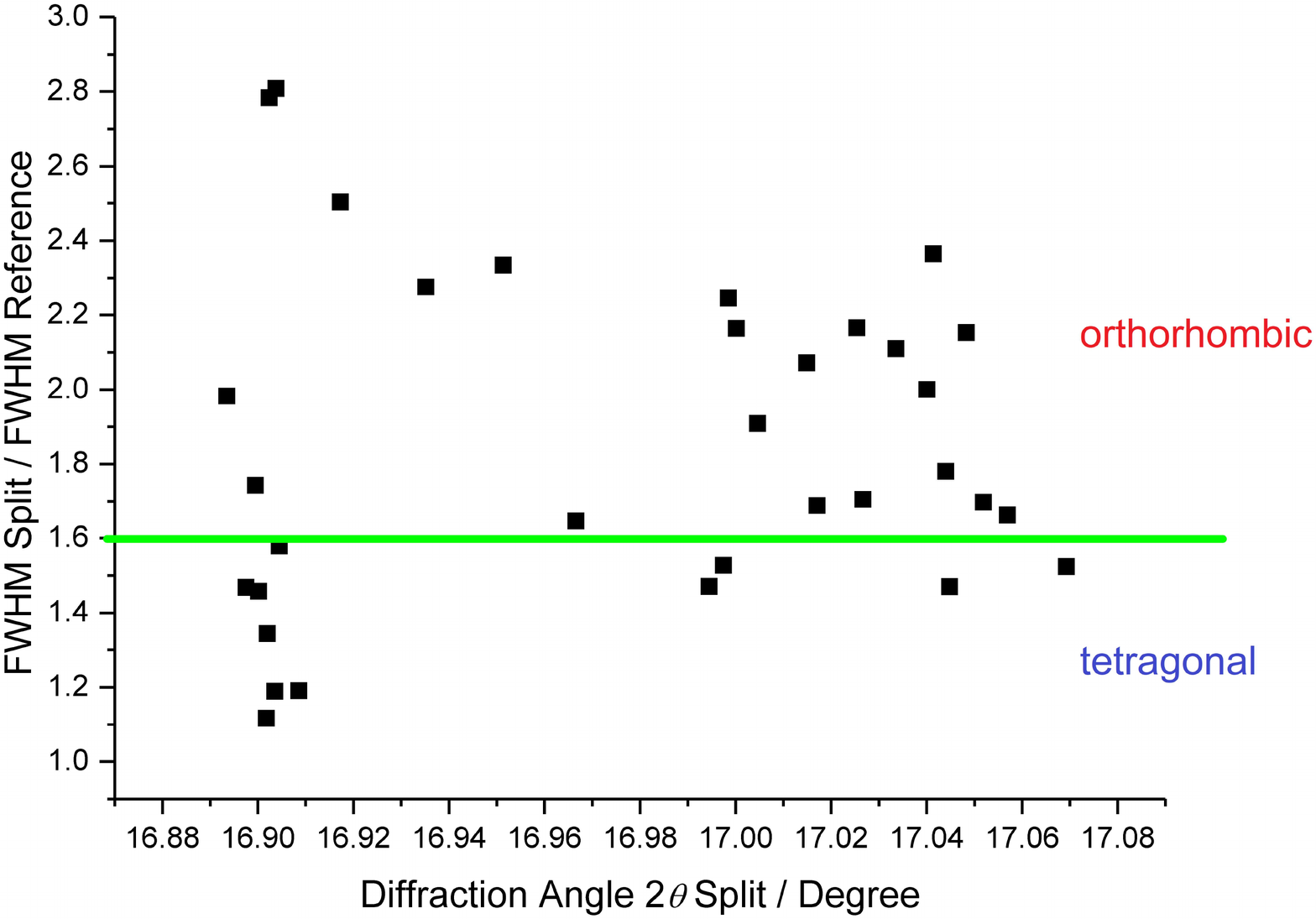}
\centering
\caption{Ratio between the FWHM of a peak which splits at the transition from the orthorhombic to the tetragonal phase and the FWHM of a reference peak showing no splitting as function of the diffraction angle of the splitting peak.
\label{criterion} }
\end{figure}

CeFeAsO displays a structural transition form a tetragonal to an orthorhombic phase upon lowering temperature. In the x-ray diffraction experiment the transition leads to a splitting of certain peaks. Figure~S\ref{diff} display an x-ray diffraction pattern at 119~K and 1.9~GPa in the upper panel. At this pressure and temperature the sample is in the orthorhombic phase. Accordingly the peak at about 17$^\circ$ is split (see inset). The lower panel of Fig.~S\ref{diff} exemplifies the evolution of  the  peak at about 17$^\circ$ and that of a reference peak which does not split upon increasing pressure at 119~K going from the orthorhombic to the tetragonal phase. There are few data points where it is hard to define, whether the sample is in the tetragonal or in the orthorhombic phase, as in the case of the data shown in Fig.\ S\ref{diff} at 2.4~GPa. Therefore, we defined a criteria to decide at which temperature/pressure point the sample is considered to be in the tetragonal or orthorhombic phase and marked that point in the temperature--pressure phase diagram in Fig.\ 3 accordingly. For the criteria we compared the FWHM of a peak which splits at the transition from the orthorhombic to the tetragonal phase and the FWHM of a reference peak which does not split. This ratio is plotted for all measured diffraction pattern in Fig.\ S\ref{criterion}. We have chosen a threshold ratio of 1.6, \textit{i.e.}\ above this value more than 60\% of the sample volume are in the orthorhombic phase.

\begin{figure*}[t!]
\includegraphics[width=0.49\linewidth]{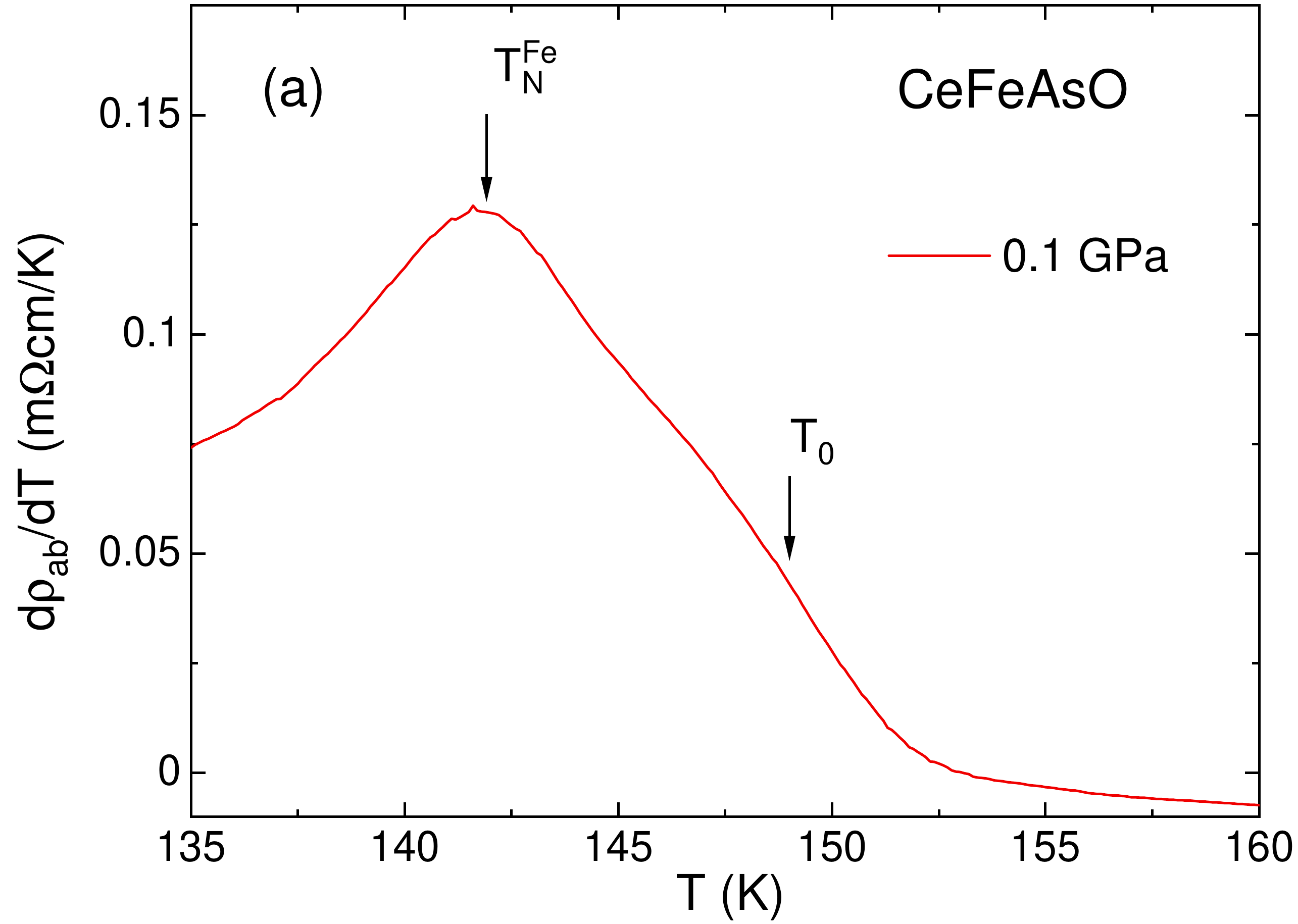}
\includegraphics[width=0.49\linewidth]{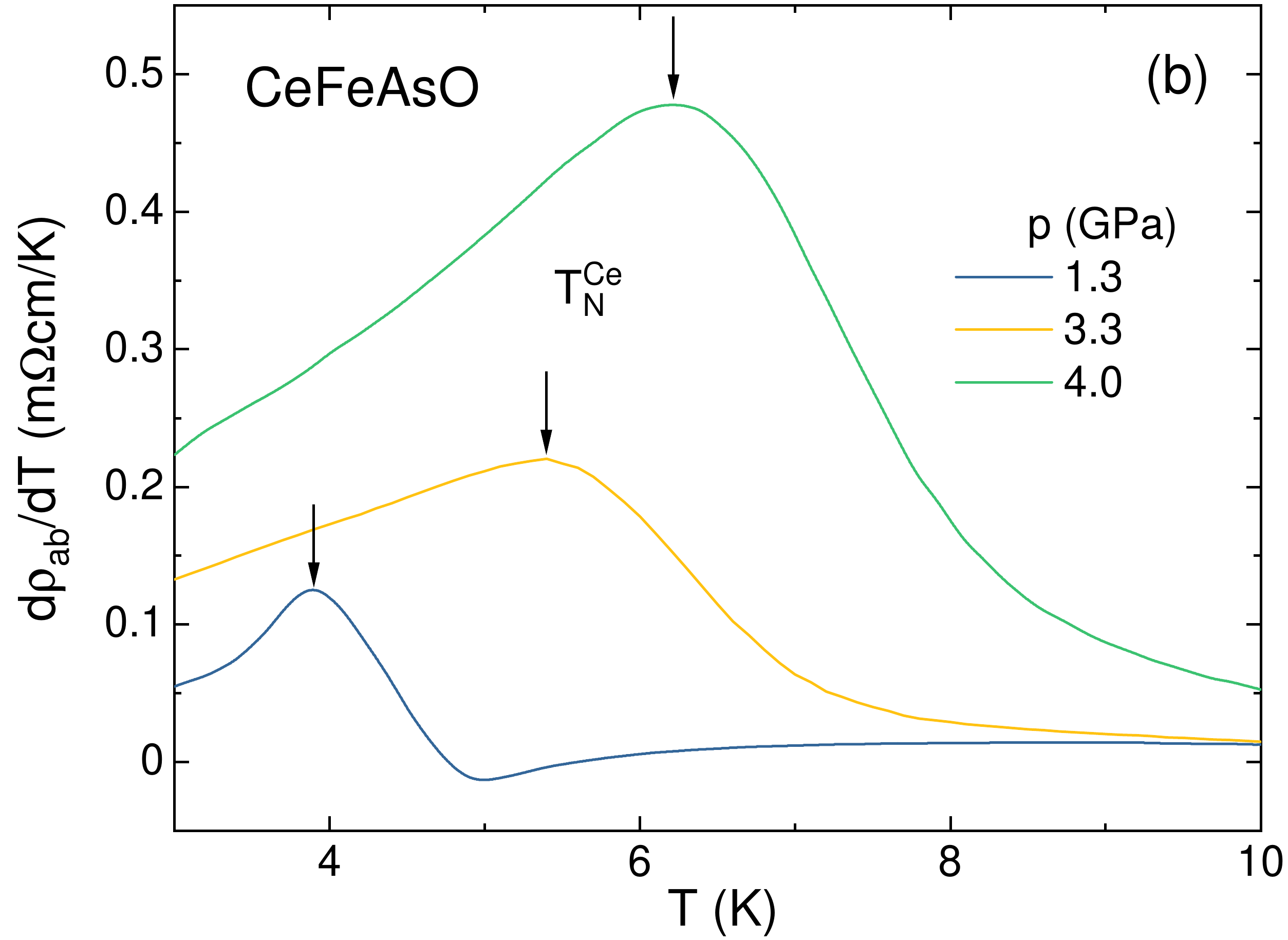}\includegraphics[width=0.49\linewidth]{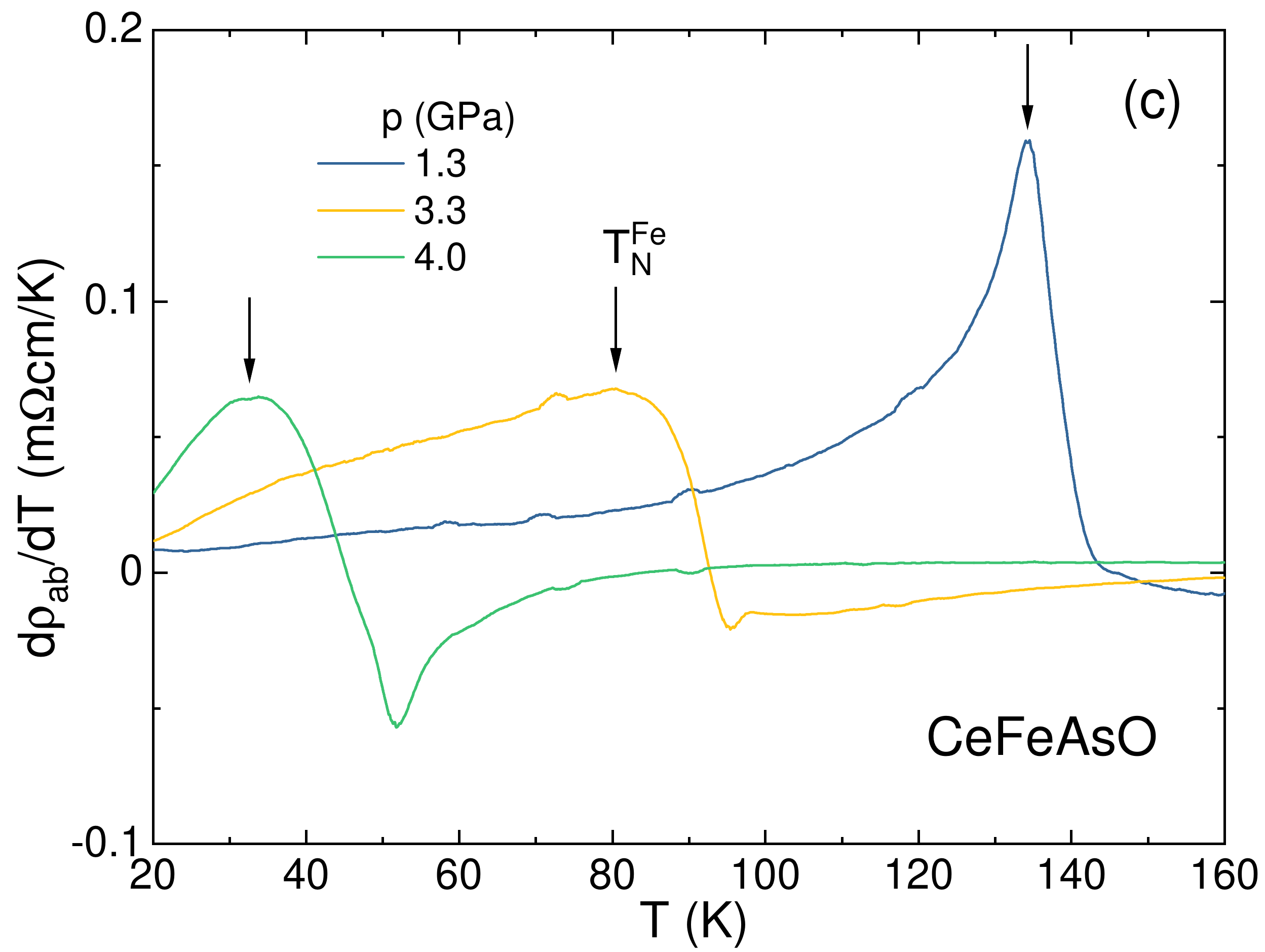}
\centering
\caption{Temperature derivative of the electrical resistivity ${\rm d}\rho_{ab}(T)/{\rm d}T$ for selected pressures in different temperature ranges. $T_N^{\rm Fe}$, $T_N^{\rm Ce}$, and $T_0$ are marked by arrows.
\label{Tcrit} }
\end{figure*}

\section*{Details on the definition of the characteristic temperatures}

$T_N^{\rm Fe} = 142$~K and $T_0=149$~K at ambient pressure are defined as maximum and inflection point in ${\rm d}\rho_{ab}(T)/{\rm d}T$ as shown in Fig.\ \ref{Tcrit}a. A second maximum in ${\rm d}\rho_{ab}(T)/{\rm d}T$ (see Fig.\ \ref{Tcrit}b) defines the ordering temperature of the Ce-$4f$ moments $T_N^{\rm Ce}$. The position of the low temperature maximum clearly shifts to higher temperatures upon increasing pressure. On the contrary, the high-temperature maximum defining $T_N^{\rm Fe}$ moves to lower temperatures upon increasing pressure (see Fig.\ \ref{Tcrit}c). The definition of the transition temperatures is the same as in Ref.\ \onlinecite{Jesche2010}.

\begin{figure}[htb!]
\includegraphics[width=0.92\linewidth]{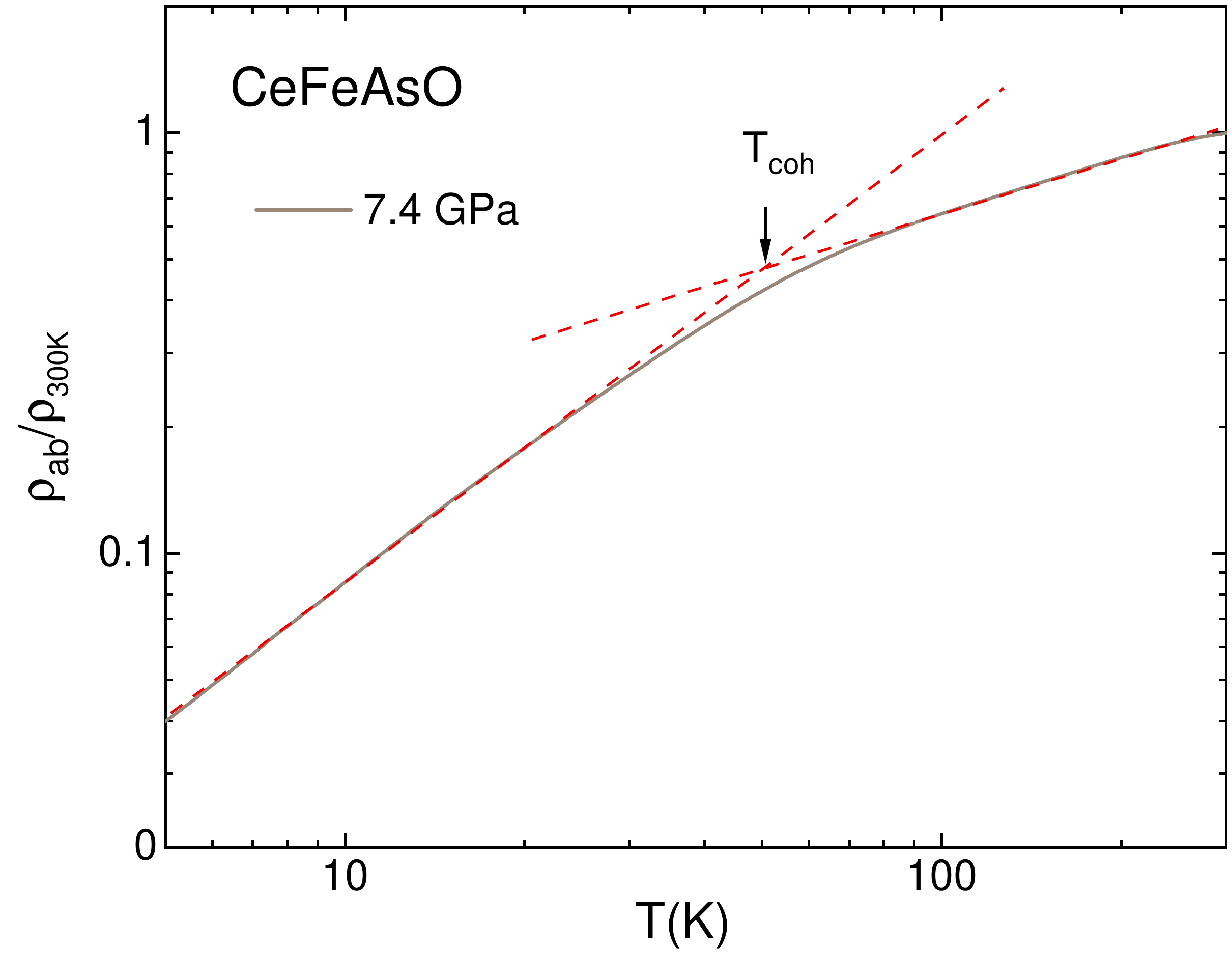}
\centering
\caption{$\rho_{ab}(T)/\rho_{\rm 300K}$ against temperature on a double-logarithmic representation for 7.4 GPa. The dashed lines are fits to the low- and high-temperature portions of the data. An arrow marks $T_{coh}$.
\label{Tcoh} }
\end{figure}
The characteristic temperature $T_{coh}$ can be estimated from the temperature below which the resistance starts to drop more rapidly. To estimate $T_{coh}$ we fit straight lines to the low- and high-temperature portions of $\rho_{ab}(T)$ on a double-logarithmic representation, as shown in Fig.\ \ref{Tcoh}, and take the position of their crossing point as $T_{coh}$. We note that this definition is only an estimate.

%

\begin{figure}[tbh!]
\includegraphics[width=0.9\linewidth]{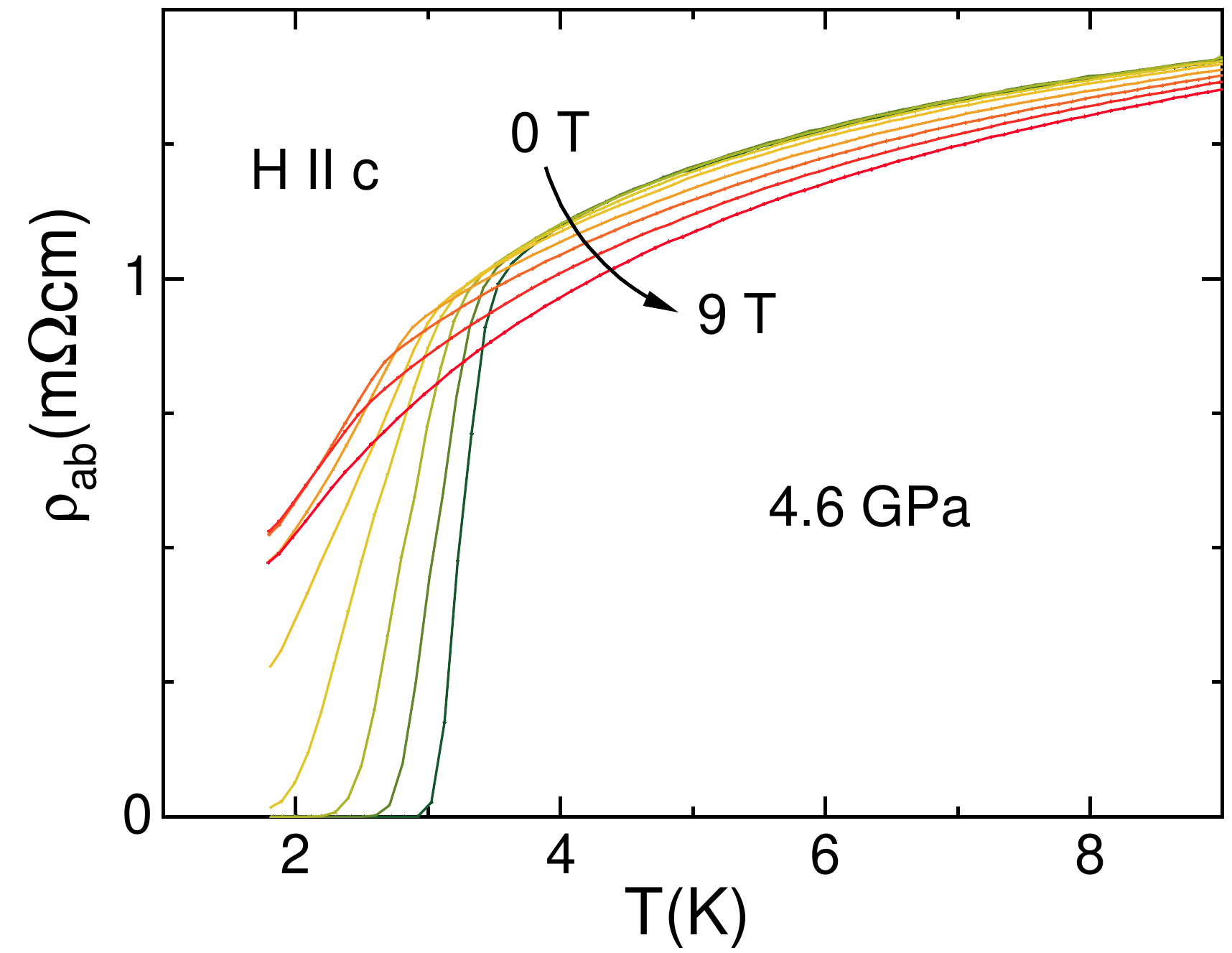}
\centering
\caption{$\rho_{ab}(T)$ at 4.6~GPa recorded in several magnetic fields between 0 and 9~T in the temperature range up to 9~K
\label{ResSC} }
\vspace{1em}
\includegraphics[width=0.9\linewidth]{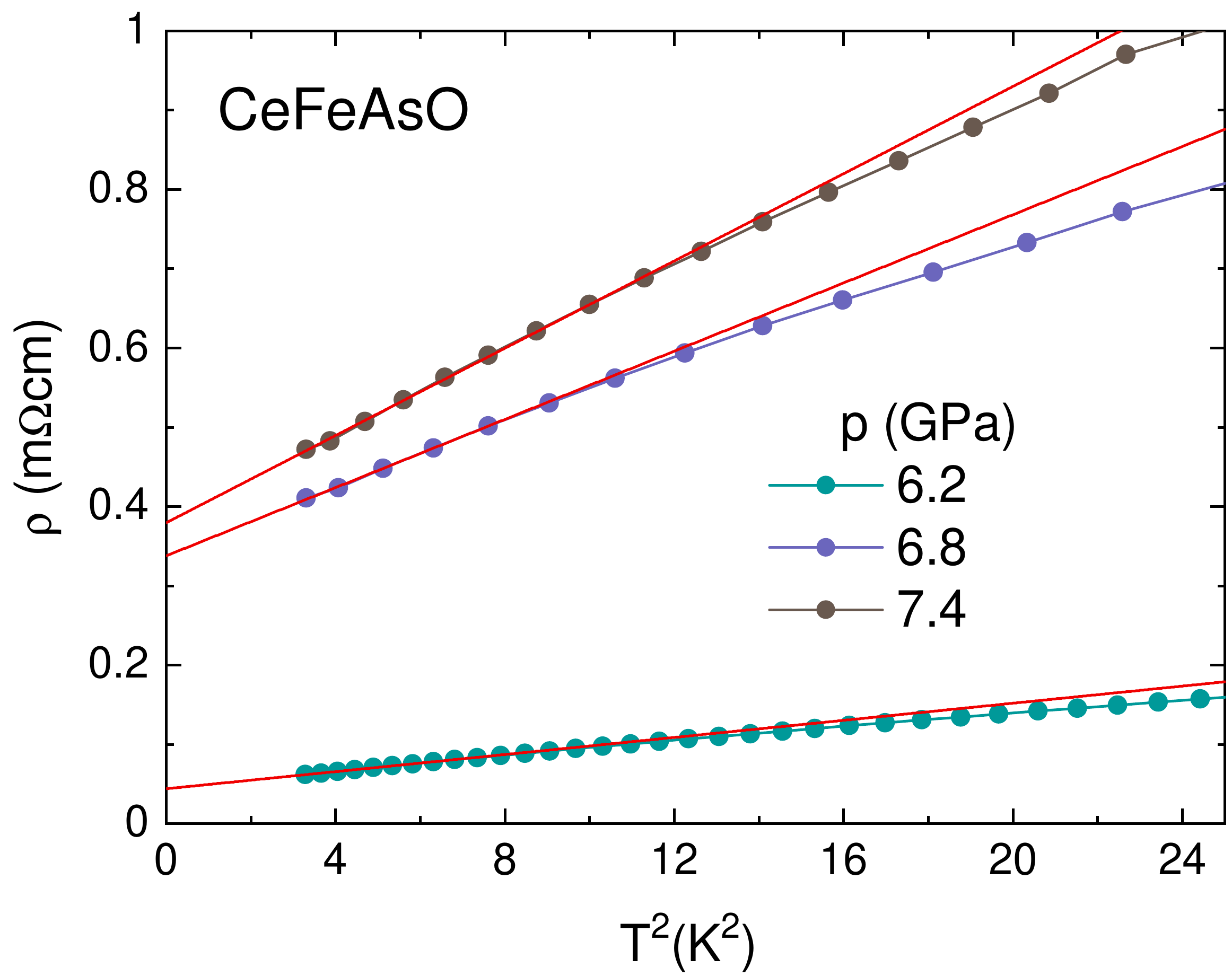}
\centering
\caption{$\rho_{ab}(T)$ at 6.2, 6.8, and 7.4~GPa plotted against $T^2$. The red lines are fits of $\rho=\rho_0+AT^2$ to the data.
\label{Acoef} }
\end{figure}

\section*{Resistivity in magnetic fields at $T_c$}


Figure S\ref{ResSC} shows resistivity data, which exemplifies how application of magnetic field suppresses the superconducting state for $H\parallel c$ in CeFeAsO at 4.6~GPa. The resistivity data has been used to establish the superconducting phase-diagram presented in Fig.\ 2e.

\section*{Low temperature resistivity}

The low-temperature resistivity data above 6.2~GPa can be described by $\rho=\rho_0+AT^2$, with the residual resistivity $\rho_0$ and the prefactor $A$. The straight lines in Fig.\ S\ref{Acoef} are fits to the data on a $T^2$ plot. It is clearly visible that the $A$ coefficient, corresponding to the slope of the straight lines, increases upon increasing pressure.\vspace{2em}

\section*{Pressure control and pressure gradient in the resistivity experiment}

In the resistivity and magnetization experiments several ruby balls of 5 to 10~micron in diameter were placed carefully next to the sample across whole length before closing the DAC. The pressure difference observed between both ends of the sample is considered as the pressure gradient across the sample. Within the resolution of our equipment ($\Delta p\approx0.05$~GPa) we do not observe any pressure distribution across the sample. A much more sensitive, but indirect way showing the good hydrostatic pressure condition across the sample comes from the obtained data. In the resistivity data on CeFeAsO we observe a sharpening and a strong temperature shift of the superconducting transition upon increasing pressure by small steps close to our resolution limit. This indicates that the pressure gradient along the sample is well below the resolution limit of our equipment ($\ll0.05$~GPa) and evidences the hydrostatic pressure conditions. Furthermore, this analysis shows the precise control of the pressure between subsequent runs, especially in the pressure range where superconductivity is observed in CeFeAsO. The control of pressure in very small steps is achieved by our precise pressurization equipment.


\end{document}